\documentclass[12pt,amsmath,amssymb,nofootinbib,superscriptaddress]{revtex4-1}
\pdfoutput=1
\usepackage{amsmath} %
\usepackage{pstricks}
\usepackage{color} 
\usepackage{amssymb}
\usepackage{slashed}
\usepackage{enumitem}
\usepackage{graphicx,amsfonts,layout,appendix,subfigure}
\input{epsf} 
\def\beq{\begin{equation}}
\def\eeq{\end{equation}}
\def\bea{\begin{eqnarray}}
\def\eea{\end{eqnarray}}
\def\beqn{\begin{eqnarray}} \def\eeqn{\end{eqnarray}}

\def\nn{\nonumber}
\def\Eq#1{Eq.~(\ref{#1})}
\def\ln#1{\mathrm{ln}\left(#1\right)}
\def\lnn#1{\mathrm{ln^2}\left(#1\right)}
\def\li#1{\mathrm{Li_2}\left(#1\right)}

\newcommand\as{a_{\mathrm{S}}}
\newcommand\alphas{\alpha_{\mathrm{S}}}

\def\beq{\begin{equation}} \def\eeq{\end{equation}}
\def\beqn{\begin{eqnarray}} \def\eeqn{\end{eqnarray}}
 \def\to{\rightarrow}
\def\nn{\nonumber}

\begin{document} 

\newcommand\sss{\scriptscriptstyle}
\newcommand\IDC{\textbf{\textit{{\rm Id}}}_C}
\newcommand\SUNT{\textit{\textbf{T}}}
\newcommand\pslashed{\slashed{p}}
\newcommand\DST{D_{\rm ST}}
\newcommand\DDirac{D_{\rm Dirac}}
\newcommand\factor{\rm Factor}
\newcommand\Spmatrix{\textit{\textbf{Sp}}}
\newcommand\Spelement{{\rm Sp}}
\newcommand\Spfunction{{\rm Split}}
\newcommand\NC{N_C}
\newcommand\CG{c_{\Gamma}}
\newcommand\DR{D_{\rm R}}
\newcommand\aem{\alpha_{\rm em}} \newcommand\refq[1]{$^{[#1]}$}
\newcommand\avr[1]{\left\langle #1 \right\rangle}
\newcommand\lambdamsb{\Lambda_5^{\rm \sss \overline{MS}}}
\newcommand\MSB{{\rm \overline{MS}}} \newcommand\MS{{\rm MS}}
\newcommand\DIG{{\rm DIS}_\gamma} \newcommand\CA{C_{\sss A}}
\newcommand\DA{D_{\sss A}} \newcommand\CF{C_{\sss F}}
\newcommand\TF{T_{\sss F}} 
\newcommand\qeps{q^2_{\epsilon}} 

\begin{titlepage}
\renewcommand{\thefootnote}{\fnsymbol{footnote}}
\begin{flushright}
     ICAS 09/16\\ IFIC/15-88
     \end{flushright}
\par \vspace{10mm}

\begin{center}
{\Large \bf
Two-loop QED corrections to the Altarelli-Parisi splitting functions
}
\end{center}

\par \vspace{2mm}
\begin{center}
{\bf Daniel de Florian}~$^{(a)}$\footnote{{\tt deflo@unsam.edu.ar}},
{\bf Germ\'an F. R. Sborlini}~$^{(b)}$\footnote{{\tt german.sborlini@ific.uv.es}}
 and 
{\bf Germ\'an Rodrigo}~$^{(b)}$\footnote{{\tt german.rodrigo@csic.es}}

\vspace{5mm}

${}^{(a)}$
International Center for Advanced Studies (ICAS), UNSAM, Campus Miguelete, 25 de Mayo y Francia,
(1650) Buenos Aires, Argentina

\vspace*{2mm}
${}^{(b)}$Instituto de F\'{\i}sica Corpuscular, 
Universitat de Val\`encia - Consejo Superior de Investigaciones Cient\'{\i}ficas,
Parc Cient\'{\i}fic, E-46980 Paterna, Valencia, Spain \\

\vspace{5mm}

\end{center}

\par \vspace{2mm}
\begin{center} {\large \bf Abstract} \end{center}
\begin{quote}
\pretolerance 10000

We compute the two-loop QED corrections to the Altarelli-Parisi (AP) splitting functions by using a
deconstructive algorithmic Abelianization of the well-known NLO QCD corrections.
We present explicit results for the full set of splitting kernels in a basis that includes the leptonic distribution functions that, starting from this order in the QED coupling, couple to the partonic densities.  
Finally, we perform a phenomenological analysis of the impact of these corrections in the splitting functions.

\end{quote}

\vspace*{\fill}
\begin{flushleft}

September 2016

\end{flushleft}
\end{titlepage}

\setcounter{footnote}{0}


\setcounter{footnote}{0}
\renewcommand{\thefootnote}{\fnsymbol{footnote}}

\section{Introduction}
\label{sec:introduccion}
The availability of highly precise experimental data collected in the LHC Run II demands to push forward the accuracy frontier at the theoretical side. In the context of perturbative QCD, higher-order corrections have been computed for a large variety of processes, reaching even N$^3$LO accuracy in some cases. In consequence,  some contributions, that were considered sub-leading long time ago, are starting to compete with QCD effects and might be crucial to compare theoretical predictions with experimental data.

Besides the substantial work accomplished in the perturbative sector, it is essential to reach the same level of accuracy on the non-perturbative side, i.e., on the parton distribution functions (PDFs). 
The calculation of the NNLO corrections to the splitting functions performed in \cite{Moch:2001im,Moch:2004pa,Vogt:2004mw,Vogt:2005dw} and the development of modern parton distribution analysis \cite{Alekhin:2013nda,Dulat:2015mca,Jimenez-Delgado:2014twa,Abramowicz:2015mha,Harland-Lang:2014zoa,Ball:2014uwa} allows to achieve the required precision in QCD.

Since $\alphas^2\sim \alpha$, NLO ElectroWeak (EW) corrections compete with NNLO QCD contributions. As a result, an accurate description for many observables requires the inclusion of the corresponding EW effects, which might account for a few percent level correction.
Recent work has been performed on the PDF sector to incorporate the EW effect (the dominant QED terms) in the evolution equations \cite{Martin:2004dh,Ball:2013hta,Bertone:2013vaa}. The appearance of the photon and the leptonic densities is the first main modification in the evolution of PDFs due to the inclusion of QED corrections. The EW corrections to PDFs need to be carefully studied for precise predictions at the LHC, as concluded from modern analysis performed up to NNLO in QCD and LO in QED. In fact, it was shown that the corrections induced are non negligible and, moreover, become crucial at higher energies \cite{Schmidt:2015zda,Sadykov:2014aua,Carrazza:2015dea,Bertone:2015lqa}.

Heretofore, the evolution of parton densities was performed using only LO QED kernels. Recently, we presented the calculation of the NLO combined QCD--QED contributions (i.e., ${\cal O}(\alpha \, \alphas)$) to the evolution kernels \cite{deFlorian:2015ujt}. Also, one-loop corrections to double~\cite{Gluck:1983mm,Gluck:1991ee,Fontannaz:1992gj,Sborlini:2013jba}
and triple~\cite{Sborlini:2014hva,Sborlini:2014mpa,Sborlini:2014kla} collinear splitting functions with photons have been computed.
With the improvement of accuracy as the main motivation, we present for the first time, the expressions for the Altarelli-Parisi (AP) splitting functions \cite{Altarelli:1977zs} to ${\cal O}(\alpha^2 )$ that completes the full set of two-loop kernels. Following the algorithmic procedure developed in Ref. \cite{deFlorian:2015ujt}, we make use of the diagram-by-diagram classification available in the original NLO QCD results presented in Refs. \cite{Curci:1980uw,Furmanski:1980cm,Ellis:1996nn} and, then, we modified consistently their color structure to account for the gluon-photon replacement. In this case, we explicitly concentrate on the QED corrections, without including those arising from Weak bosons, which only become relevant for very extreme kinematical conditions, where their masses are neglected in comparison to other scales involved in the process.

The structure of the paper is as follows. In Section \ref{sec:kerneldefinitions} we recall the evolution equations for the different distributions and the corresponding kernels, introducing the notation required to present our results. Also, we present there the constraints from sum rules that determine the behaviour of splitting kernels in the end-point region (i.e. $x=1$). In Section \ref{sec:results} we summarize the algorithm that we use to obtain the QED corrections to the splitting functions and present the corresponding kernels. Using these formulae, we study the changes introduced in the AP kernels by both ${\cal O}(\alpha^2)$ and ${\cal O}(\alpha \, \alphas)$, in Section \ref{sec:pheno}. Finally, conclusions are given in Section \ref{sec:conclusions}.

\section{Evolution equations and definitions}
\label{sec:kerneldefinitions}
In the context of combined QCD--QED contributions, it is mandatory to take into account lepton distributions. In Ref.~\cite{deFlorian:2015ujt}, we computed the ${\cal O}(\alpha \, \alphas)$ contributions to the AP kernels and we showed that leptons decouple from the QCD sector at that accuracy. Thus, in that case, we neglected lepton distributions. Moreover, this simplification remains true for ${\cal O}(\alpha \, \alphas^n)$ because the  quark-lepton mixing appears starting at ${\cal O}(\alpha^2)$. Therefore, here we follow the path established in Ref. \cite{Roth:2004ti} and obtain the exact set of evolution equations for the combined QCD--QED model in a proper basis.

As usual, the first step consists in writing down the evolution equations for quark, lepton, gluon and photon distributions. These equations are obtained starting from those available in Ref.\cite{deFlorian:2015ujt} by adding lepton distributions and the corresponding AP kernels, $P_{ij}$. Using the standard definition for the convolution operator, i.e.
\beq
(f\otimes g)(x) = \int_x^1 \, \frac{dy}{y} \, f\left(\frac{x}{y}\right)g(y) \, ,
\eeq 
and introducing $t=\ln{\mu^2}$ as the evolution variable, we have
\beqn
\frac{dg}{dt}= \sum_{f} P_{g f} \otimes  f + \sum_{f} P_{g \bar{f}} \otimes  \bar{f} +  P_{g g} \otimes g +  P_{g \gamma} \otimes \gamma \, ,
\label{eq:evoluciongluon}
\\ \frac{d\gamma}{dt}= \sum_{f} P_{\gamma f} \otimes  f + \sum_{f} P_{\gamma \bar{f}} \otimes  \bar{f} +  P_{\gamma g} \otimes g +  P_{\gamma \gamma} \otimes \gamma \, ,
\label{eq:evolucionphoton}
\\ \frac{dq_i}{dt}= \sum_{f} P_{q_i f} \otimes f + \sum_{f} P_{q_i \bar{f}} \otimes \bar{f} +  P_{q_i g} \otimes g +  P_{q_i \gamma} \otimes \gamma \, ,
\\ \frac{dl_i}{dt}= \sum_{f} P_{l_i f} \otimes f + \sum_{f} P_{l_i \bar{f}} \otimes \bar{f} +  P_{l_i g} \otimes g +  P_{l_i \gamma} \otimes \gamma \, ,
\label{eq:evolucionlepton}
\eeqn
and similarly for antiparticles by using charge conjugation invariance. Here the sum over fermions $f$ runs over all the active flavours of quarks ($n_F$) and leptons ($n_L$). In the previous formulae, $\mu$ represents the factorization scale.

Along this work we will present the expressions for the splitting functions including QCD and QED corrections. Thus, we expand them according to
\beqn
P_{ij} = \as \, P_{ij}^{(1,0)} +a \, P_{ij}^{(0,1)} +\as^2 \, P_{ij}^{(2,0)} + \as \, a \, P_{ij}^{(1,1)}+a^2 \, P_{ij}^{(0,2)} + ... \, ,
\label{eq:expansionkernels}
\eeqn
where the upper indices indicate the (QCD,QED) order of the calculation, while
\beq
a_{\mathrm{S}}\equiv \frac{\alpha_{\mathrm{S}}}{2\pi} \ \ \ , \ \ \ a \equiv \frac{\alpha}{2\pi} \, ,
\eeq
allow to set the standard normalization of the splitting functions. 

The presence of electromagnetic interactions introduces a charge dependence in the splitting functions. Moreover, due to higher-order QED corrections, a mixing among leptons and QCD partons might take place, which leads to more complicated evolution equations. In fact, in the most general case, Eqs.~(\ref{eq:evoluciongluon})-(\ref{eq:evolucionlepton}) constitute a system of $20\times 20$ coupled first-order differential equations. However, notable simplifications are achieved at each order of the truncated expansion by imposing physical constraints. For instance, at ${\cal O}(\alpha \, \alphas^n)$, the kernels depend on the electric charge of the initiating fermions (up or down), such that in general $P_{q}^{(1,n)} \sim e_q^2$. As we will show later, at ${\cal O}(\alpha^2)$, the charge content of the  $P_{q l}$ kernels becomes non trivial due to the exchange of a pair of  photons.

The quark splitting functions are decomposed as
\beqn
P_{q_i \, q_k} &=& \delta_{ik} \, P^V_{qq} + P^S_{qq} \, ,
\\ P_{q_i \, \bar{q}_k} &=& \delta_{ik} \, P^V_{q\bar{q}} + P^S_{q\bar{q}} \, ,
\\ P_q^{\pm} &=& P^V_{qq} \pm P^V_{q\bar{q}} \, ,
\eeqn
which act as a definition of $P^V_{q q}$ and $P^V_{q \bar{q}}$, i.e. the non-singlet components. In a completely analogous way, we write
\beqn
P_{l_i \, l_k} &=& \delta_{ik} \, P^V_{ll} + P^S_{ll} \, ,
\\ P_{l_i \, \bar{l}_k} &=& \delta_{ik} \, P^V_{l\bar{l}} + P^S_{l\bar{l}} \, ,
\\ P_l^{\pm} &=& P^V_{ll} \pm P^V_{l\bar{l}} \, ,
\eeqn
for the lepton kernels. For mixed lepton-quark splittings we use $P^S_{l q}\equiv P_{l q}$ to simplify the notation.

The canonical basis of distributions is given by
\beq
{\cal B}_{\rm c} = \{u,\bar{u},\ldots ,t,\bar{t},e,\bar{e}, \ldots,  \tau,\bar{\tau},g,\gamma \} \, ,
\eeq
when considering the full charged-fermion content of the Standard Model. Working with a reduced amount of fermions simply involves removing them from the previous basis. As explained in Ref.\cite{Roth:2004ti}, ${\cal B}_{\rm c}$ is not the optimal choice to reduce the mixing among the different parton distributions in the evolution system. Thus, for $n_F=5$ active flavours it is more suitable to work with the following set:
\beq
{\cal B} = \{u_v,d_v,s_v,c_v,b_v,e_v,\mu_v,\tau_v,\Delta_{uc},\Delta_{ds},\Delta_{sb},\Delta_2^l,\Delta_{UD}, \Delta_3^l,\Sigma,\Sigma^l,g,\gamma \} \, ,
\eeq
where
\beqn
q_v &=& q_i-\bar{q_i} \, , \\
l_v &=& l_i-\bar{l_i} \, ,
\eeqn
are the valence distributions, whilst
\beqn
\Delta_{uc} &=& u+\bar{u}-c-\bar{c} \, ,  \\ 
\Delta_{ds} &=& d+\bar{d}-s-\bar{s} \, , \\ 
\Delta_{sb} &=& s+\bar{s}-b-\bar{b} \, , \\ 
\Delta_{2}^l &=& e+\bar{e}-\mu-\bar{\mu} \, , \\ 
\Delta_{UD} &=& u+\bar{u}+c+\bar{c} -d-\bar{d} -s-\bar{s}-b-\bar{b} \, ,\\
\Delta_{3}^l &=& e+\bar{e}+\mu+\bar{\mu} -2(\tau+\bar{\tau}) \, , \\ 
 \Sigma &=& \sum_{i=1}^{n_F} ( q_i+\bar{q}_i) \, , \\
 \Sigma^l &=& \sum_{i=1}^{n_L} ( l_i+\bar{l}_i) \, ,  
\eeqn
are the remaining combinations (gluons and photons are treated separately). To include the top distribution in case of a six flavour analysis, it is necessary to introduce the elements $\{t_v,\Delta_{ct}\}$ and to extend the definitions of $\Delta_{UD}$ and $\Sigma$.

As we mentioned before, QED interactions introduce a classification of the different fermions according to the absolute value of their electromagnetic (EM) charges. Thus, there are three possible fermionic sectors: up-like quarks ($u$ and $e_u=2/3$), down-like quarks ($d$ and $e_d=1/3$) and leptons ($l$ with $e_l=1$). Particles inside each sector are indistinguishable by QCD--QED interactions. Also, it is useful to define 
\beqn
\Delta P_{f F}^S & \equiv & P^S_{f F} -P^S_{f \bar{F}} \,  ,
\\ \bar{P}_{f F}^S & \equiv &  P^S_{f F} + P^S_{f \bar{F}} \, , 
\eeqn
where $f$ and $F$ denote the possible fermion subgroups ($u$, $d$ or $l$). 
Notice that in the context of QCD--QED, it might occur that $P_{l q} \neq P_{q l}$ due to higher-order contributions. However, at ${\cal O}(\alpha^2)$, they are the same and the equality can be used to achieve further simplifications. Moreover, at this order, it is verified that 
\beq
\Delta P_{fF}^S  \equiv 0 \, ,
\label{eq:simplificacionnoQQBAR}
\eeq
due to charge conjugation invariance.

In the most general case, the corresponding QCD--QED combined evolution equations for the distributions in the optimized basis are given by:
 \beqn
\frac{dq_{v_i}}{dt} &=&   P_{q_i}^-     \otimes q_{v_i}  +\sum_{j=1}^{n_F} \Delta P_{q_i q_j}^S  \otimes  q_{v_j} + \Delta P_{q_i l}^S \otimes \left(\sum_{j=1}^{n_L}   l_{v_j} \right)    \, ,
\label{eq:evolucionqvGEN}
\\ \frac{dl_{v_i}}{dt} &=&   P_{l}^-     \otimes l_{v_i}  +\sum_{j=1}^{n_F} \Delta P_{l q_j}^S  \otimes  q_{v_j} + \Delta P_{ll}^S \otimes \left(\sum_{j=1}^{n_L}   l_{v_j} \right)    \, ,
\label{eq:evolucionlvGEN}
\eeqn
for valence distributions,
\beqn
\nn \frac{d  \Sigma  }{dt} &=&  \frac{P_{u}^+ + P_{d}^+}{2} \otimes \Sigma
                                                  + \frac{P_{u}^+ - P_{d}^+}{2} \otimes \Delta_{UD} + \frac{n_u \bar{P}^S_{uu} +n_d \bar{P}^S_{dd} + (n_u+n_d) \bar{P}^S_{ud}}{2} \, \otimes \Sigma
\\ \nn &+& \frac{n_u \bar{P}^S_{uu} -n_d \bar{P}^S_{dd} - (n_u-n_d) \bar{P}^S_{ud}}{2} \, \otimes \Delta_{UD} + \left(n_u \bar{P}^S_{ul}+n_d \bar{P}^S_{dl} \right) \, \otimes \Sigma^l
                                                  \\ &+& 2 (n_u P_{ug} +n_d P_{dg})  \otimes g 
                                                  + 2 (n_u P_{u\gamma} +n_d P_{d\gamma})  \otimes \gamma \, ,
                                                  \label{eq:evolucionSIGMAGEN}
                                                  \\ \nn \frac{d  \Sigma^l  }{dt} &=& n_L \frac{\bar{P}^S_{lu}+\bar{P}^S_{ld}}{2} \otimes \Sigma
                                                  + n_L \frac{\bar{P}^S_{lu}- \bar{P}^S_{ld}}{2} \otimes \Delta_{UD} +  \left(P_l^+ + n_L \bar{P}^S_{ll} \right) \otimes \Sigma^l
                                                  \\ &+& 2 n_L (P_{lg} \otimes g + P_{l \gamma}   \otimes \gamma) \, ,
                                                  \label{eq:evolucionSIGMAlGEN}
\eeqn
for the singlets and
\beqn
\frac{d \{ \Delta_{uc} , \Delta_{ct} \}}{dt} &=&  P_{u}^+ \otimes \{ \Delta_{uc} , \Delta_{ct}  \} \, ,
\label{eq:evolucionDupperGEN}
\\ \frac{d \{ \Delta_{ds} ,\Delta_{sb}  \}}{dt} &=&  P_{d}^+ \otimes \{ \Delta_{ds} ,\Delta_{sb}  \} \, ,
\label{eq:evolucionDlowerGEN}
\\ \frac{d \Delta_{2}^l }{dt} &=&  P_{l}^+ \otimes \Delta_{2}^l \, ,
\label{eq:evolucionD2leptonGEN}
\eeqn
\beqn
\nn \frac{d  \Delta_{UD}  }{dt} &=&  \frac{P_{u}^+ + P_{d}^+}{2} \otimes \Delta_{UD} 
                                                  + \frac{P_{u}^+ - P_{d}^+}{2} \otimes \Sigma +  \frac{n_u \bar{P}^S_{uu} -n_d \bar{P}^S_{dd} + (n_u-n_d) \bar{P}^S_{ud}}{2} \, \otimes \Sigma
\\ \nn &+& \frac{n_u \bar{P}^S_{uu} +n_d \bar{P}^S_{dd} - (n_u+n_d) \bar{P}^S_{ud}}{2} \, \otimes \Delta_{UD} + \left(n_u \bar{P}^S_{ul}-n_d \bar{P}^S_{dl} \right) \, \otimes \Sigma^l                                                   
                                                  \\ &+& 2 (n_u P_{ug} -n_d P_{dg})  \otimes g 
                                                  + 2 (n_u P_{u\gamma} -n_d P_{d\gamma})  \otimes \gamma \, ,
                                                  \label{eq:evolucionDUDGEN}
\\ \frac{d \Delta_{3}^l }{dt} &=&  P_{l}^+ \otimes \Delta_{3}^l \, ,
\label{eq:evolucionD3leptonGEN}
\eeqn
for the remaining fermionic distributions. Here $n_u (n_d)$ refers to the active number of $u(d)$ type-quarks, respectively ($n_F=n_u+n_d$), and $n_L$ is the number of leptons under consideration. It is worth noticing that only $\Delta_{ij}$ and $\Delta_i^l$ decouple from the other distributions. Besides that, if we restrict to ${\cal O}(\alpha \, \alphas)$, we recover the equations presented in Ref. \cite{deFlorian:2015ujt}. At ${\cal O}(\alpha^2)$, the splitting kernels are charge dependent but $\Delta P^S \equiv 0$. Thus, in that case,  Eqs. (\ref{eq:evolucionqvGEN}) and (\ref{eq:evolucionlvGEN}) become
\beqn
\frac{dq_{v_i}}{dt} &=&   P_{q_i}^-     \otimes q_{v_i}    \, ,
\label{eq:evolucionqvSIMPLE}
\\ \frac{dl_{v_i}}{dt} &=&   P_{l}^-     \otimes l_{v_i}    \, ,
\label{eq:evolucionlvSIMPLE}
\eeqn
and
\beq
\bar{P}_{ij}^S \equiv 2 P^S_{ij}\, , \ \ \ \ P^S_{lq} \equiv P^S_{ql} \, \, .
\eeq
Moreover, if we only allow QED interactions, all splitting kernels with gluons vanish and the gluon distribution is decoupled from the other ones.

\subsection{Constraints from Sum Rules}
\label{ssec:constraints}
On one side, QCD--QED interactions preserve the fermion number. In particular, this implies that splitting kernels must fulfil
\beqn
\int_0^1 dx \, P^{-}_f &=& 0 \, ,
\label{eq:constraintfermionumber}
\eeqn
because the factorization scale $\mu$ is arbitrary. On the other hand, the arbitrariness of $\mu$ also implies that the momentum of the proton is conserved during the evolution. Using the parton model, we have
\beqn
0 = \frac{d P}{dt} &=& \int_0^1 dx \, x \, \left(\frac{dg}{dt}+\frac{d\gamma}{dt}+\sum_f \left(\frac{df}{dt}+\frac{d\bar{f}}{dt} \right) \right)  \, ,
\label{eq:protonmomentum}
\eeqn
where the sum is over all the possible fermion flavours (both quarks and leptons). If we express \Eq{eq:protonmomentum} by using the optimized basis, we impose its validity in each component. The non-trivial constraints are:
\begin{itemize}
\item Gluon and photon components,
\beqn
&& \int_0^1 dx \, x \, \left(2 n_d P_{dg} + 2 n_u P_{ug} +2 n_L P_{lg} + P_{\gamma g} + P_{gg} \right) = 0\, ,
\label{eq:constraintgluon}
\\ && \int_0^1 dx \, x \, \left(2 n_d P_{d \gamma} + 2 n_u P_{u \gamma} +2 n_L P_{l \gamma} + P_{g\gamma} + P_{\gamma\gamma} \right) = 0 \, ;
\label{eq:constraintgamma}
\eeqn
\item $\Delta_{UD}$ component,
\beqn
\nn \int_0^1 dx \, x && \, \left(\frac{P^+_u-P^+_d}{2}+n_L \frac{\bar{P}^S_{lu}-\bar{P}^S_{ld}}{2}+\frac{n_u \bar{P}^S_{uu}-n_d \bar{P}^S_{dd}}{2} -\frac{(n_u-n_d) \bar{P}^S_{ud}}{2} \right.
\\ && \left. + \frac{P_{gu}-P_{gd}}{2}+\frac{P_{\gamma u}-P_{\gamma d}}{2} \right) = 0\, , \ \ 
\label{eq:constraintDelta}
\eeqn
\item and, finally, the singlet components ($\Sigma$ and $\Sigma^l$, respectively),
\beqn
\nn \int_0^1 dx \, x && \, \left(\frac{P^+_u+P^+_d}{2}+n_L \frac{\bar{P}^S_{lu}+\bar{P}^S_{ld}}{2}+\frac{n_u \bar{P}^S_{uu}+n_d \bar{P}^S_{dd}}{2} + \frac{n_F \bar{P}^S_{ud}}{2} \right. 
\\ && \left. + \frac{P_{gu}+P_{gd}}{2}+\frac{P_{\gamma u}+P_{\gamma d}}{2} \right) = 0\, , \ \ \
\label{eq:constraintSigma}
\\ \int_0^1 dx \, x && \, \left(n_u \bar{P}^S_{u l} + n_d \bar{P}^S_{d l} + n_L \bar{P}^S_{ll} +P_l^+ + P_{gl} +P_{\gamma l} \right) = 0 \, . \ \ \
\label{eq:constraintSigmaL}
\eeqn
\end{itemize}
In the following sections, we will use these equations to provide a strict check of the calculation and, at the same time, fix the value of the splitting kernels in the end-point $x=1$.

\section{Splitting kernels at ${\cal O}(\alpha^2)$}
\label{sec:results}
Let's start by recalling some well-known results for the lowest order splitting functions. At ${\cal O}(\alphas)$, only QCD partons are involved \cite{Altarelli:1977zs}; thus,
\beqn
P_{qq}^{(1,0)}(x) &=& C_F \left[  \frac{1+x^2}{(1-x)_+}  +\frac{3}{2} \, \delta (1-x) \right] = 
C_F \left[ \, p_{qq}(x) + \frac{3}{2} \, \delta(1-x)\right] \, , \nn \\
P_{qg}^{(1,0)}(x) &=& T_R \left[   x^2+(1-x)^2\right]= T_R \, p_{qg}(x) \, ,\nn \\
P_{gq}^{(1,0)}(x) &=& C_F \left[   \frac{1+(1-x)^2}{x}  \right]= C_F \, p_{gq}(x) \, , \nn \\
P_{gg}^{(1,0)}(x) &=& 2 \, C_A \left[  \frac{x}{(1-x)_+}   +\frac{1-x}{x} + x(1-x) \right] +  \frac{\beta_0}{2} \, \delta(1-x) \, ,
\eeqn
with $\beta_0=\frac{11 N_C-4 n_F T_R}{3}$ and the plus distribution defined as
\beq
\int_0^1 dx \, \frac{f(x)}{(1-x)_+} = \int_0^1 dx \, \frac{f(x)-f(1)}{1-x} \, ,
\eeq
for any regular test function $f$. As usual, the normalization of the fundamental representation is set to $T_R=1/2$ and
\beq
C_A = N_C \, , \quad \quad C_F= \frac{N_C^2-1}{2\, N_C} \, , 
\eeq
are the ${\rm SU}(N_C)$ group factors. In particular, for QCD ($N_C=3$), we have $C_A=3$ and $C_F=4/3$. Notice that these expressions provide a definition for the color-stripped splitting functions $p_{ij}$, which will be used to simplify the presentation of higher-order corrections. At ${\cal O}(\alpha)$, splitting processes can be described by replacing the color factors in $P_{ij}^{(1,0)}$ with the corresponding EM charges. In this way, we have \cite{Roth:2004ti}
\beqn
P_{ff}^{(0,1)}(x) &=& e_f^2 \left[  p_{qq}(x) +\frac{3}{2} \delta (1-x) \right] \, , \nn \\
P_{f\gamma}^{(0,1)}(x) &=& e_f^2  \, p_{qg}(x) \, ,  \nn \\
P_{\gamma f}^{(0,1)}(x) &=&  e_f^2 \, p_{gq}(x) \, , \nn \\
P_{\gamma\gamma}^{(0,1)}(x) &=& -\frac{2}{3} \sum_{f} e_f^2 \, \delta (1-x)  \, ,
\eeqn
where $f$ denotes any fermion (quark or lepton) with its corresponding EM charge $e_f$, and
\beq
\sum_{f}\, e_f^a =N_C \sum_{j=1}^{n_F} \, e_{q_j}^a \, + \, \sum_{j=1}^{n_L} \, e_{l_j}^a  \, ,
\eeq
is the sum over fermion charges, taking into account that quark-photon interactions are degenerate due to color degrees of freedom ($N_C$). Also, in the case of $P^{(0,1)}_{f\gamma}$ an extra factor of $N_C$ has to be included whenever the fermion $f$ is a colored quark.

In order to obtain the pure two-loop QED corrections $P_{ij}^{(0,2)}$, we follow the ideas depicted in Ref. \cite{deFlorian:2015ujt}. We start from the results on the two-loop QCD anomalous dimensions in the light-cone gauge, originally performed for the non-singlet component by Curci, Furmanski and Petronzio in Ref. \cite{Curci:1980uw} and extended to the singlet case in Refs. \cite{Furmanski:1980cm,Ellis:1996nn}\footnote{For a complete review of the previous developments in the computation of higher-order corrections to the splitting kernels and the anomalous dimension, see Ref. \cite{Blumlein:2012bf} and the references therein.}. Then we take the corresponding Abelian limit, which involves replacing each gluon by a photon \cite{Kataev:1992dg}. This automatically avoids the presence of diagrams with non-Abelian vertices (at least in pure QED). The last step consists in replacing the original color structure with the one obtained after the double replacement $g \to \gamma$, and multiplying by the EM charge of the fermions involved in the process.

However, as anticipated in Sec. \ref{sec:kerneldefinitions}, lepton distributions enter in the evolution of the system at ${\cal O}(\alpha^2)$, which forces us to compute also lepton-quark and lepton-photon kernels at this order. The procedure is completely analogous to the one described before.

There is a subtlety related with the presence of quark loops in pure QCD results. In that case, gluons couple in the same way to all quark flavours, originating a factor $n_F$. Once we replace gluons with photons, virtual leptons are also allowed inside the loop. Both for leptons and quarks, the QED coupling is proportional to their EM charges. In consequence, the replacement
\beq
n_F \to \sum_{f} \, e_f^2 \, ,
\eeq
has to be implemented in all the contributions arising from quark loops in the pure QCD kernels.

Another subtle point that we must carefully treat is the presence of massive EW bosons. As we mentioned before, we neglect their contribution in this work. This is due to the fact that their mass is kept strictly non-vanishing, thus acting as an IR-regulator. In other terms, IR-singular diagrams for processes involving heavy EW bosons can always be treated by making use of QCD--QED splitting functions and factorizing the massive particle into the hard scattering subprocess.

So, let's present the explicit results. In first place, kernels involving gluons vanish at this order; hence,
 \beqn
\nn && P_{fg}^{(0,2)} = 0  \, , \ \ \ \ P_{gf}^{(0,2)} = 0  \, , \ \ \ \ P_{\gamma g}^{(0,2)} = 0 \, ,
\\ && P_{g \gamma}^{(0,2)} = 0 \, , \ \ \ \ P_{gg}^{(0,2)} = 0  \, .
\eeqn
Then, we consider those kernels which involve quarks and photons,
\beqn
\nn P_{q\gamma}^{(0,2)} &=& \frac{C_A \, e_q^4}{2} \left\{ \vphantom{2\lnn{\frac{1-x}{x}}} 4 -9x-(1-4x) \ln{x} -(1-2x)\lnn{x} +4 \ln{1-x} \right. 
\\ &+& \left. p_{qg}(x) \left[  2\lnn{\frac{1-x}{x}} -4\ln{\frac{1-x}{x}}  -\frac{2\pi^2}{3}+10 \right] \right\} \, ,
\\ \nn P_{\gamma q}^{(0,2)} &=& e_q^4 \left[-\left(3\ln{1-x}+\lnn{1-x}\right)p_{gq}(x)+\left(2+\frac{7}{2}x\right)\ln{x}-\left(1-\frac{x}{2}\right)\lnn{x} \right.
\\ &-& \left. 2x\ln{1-x}-\frac{7}{2}x-\frac{5}{2}\right] - e_q^2  \left(\sum_{f}\, e_f^2 \right) \left[  \frac{4}{3}x + p_{gq}(x) \left( \frac{20}{9}+\frac{4}{3}\ln{1-x} \right)\right] \, ,
\\ \nn P_{qq}^{V(0,2)} &=&  -  e_q^4 \left[\left(2 \ln{x}\ln{1-x}+\frac{3}{2}\ln{x}\right) p_{qq}(x) + \frac{3+7x}{2}\ln{x} \right. 
\\ \nn &+& \left.  \frac{1+x}{2}{\lnn{x}}+5(1-x) +\left(\frac{\pi^2}{2}-\frac{3}{8}-6 \zeta_3 \right) \delta(1-x)  \right]
\\ &-& e_q^2  \left(\sum_{f}\,e_f^2\right) \left[  \frac{4}{3}(1-x) +  p_{qq}(x) \left( \frac{2}{3} \ln{x}+\frac{10}{9}  \right) +\left(\frac{2 \pi^2}{9}+\frac{1}{6}\right) \delta(1-x)\right] \, ,
\\ P_{q\bar{q}}^{V(0,2)} &=& e_q^4 \left[4(1-x)+2(1+x)\ln{x}+2p_{qq}(-x)S_2(x)\right] \, ,
\\ P_{qQ}^{S(0,2)} &=& P_{q\bar{Q}}^{S(0,2)}= C_A \, e_q^2 \, e_Q^2 \, p_s(x)  \, ,
\eeqn 
where $\{q,Q\}$ denote different quark flavours and we defined the function
\beqn
p_s(x) &=& \frac{20}{9x}
-2+6x-\frac{56}{9}x^2+\left(1+5x+\frac{8}{3}x^2\right)\ln{x}-(1+x)\lnn{x} \, ,
\eeqn
which appears in all the higher-order corrections to the singlet components. The function $S_2(x)$ is given by \cite{Furmanski:1980cm,Blumlein:1998if}
\beqn
\nn S_2(x) &=& \int_{\frac{x}{1+x}}^{\frac{1}{1+x}} \, \frac{dz}{z} \, \ln{\frac{1-z}{z}} = \frac{\lnn{x}}{2}-\zeta_2-2 \li{-x}-2\ln{x}\ln{1+x} \, .
\label{eq:S2definicion}
\eeqn
In these formulae, $\zeta_n$ is the Riemann zeta function, which verifies $\zeta_2=\pi^2/6$ and $\zeta_3\approx 1.202057$.

In an analogous way, splitting functions with leptons and photons are given by
\beqn
P_{l\gamma}^{(0,2)} &=&  \frac{e_l^4}{C_A \, e_q^4}  \, P_{q \gamma}^{(0,2)} \, ,
\\ \nn P_{\gamma l}^{(0,2)} &=&  e_l^4 \, \left[-(3\ln{1-x}+\lnn{1-x})p_{gq}(x)+\left(2+\frac{7}{2}x\right)\ln{x}-\left(1-\frac{x}{2}\right)\lnn{x} \right.
\\ &-& \left. 2x\ln{1-x}-\frac{7}{2}x-\frac{5}{2}\right] - e_l^2  \left(\sum_{f}\, e_f^2\right) \left[  \frac{4}{3}x + p_{gq}(x) \left( \frac{20}{9}+\frac{4}{3}\ln{1-x} \right) \right] \, ,
\\ \nn P_{ll}^{V(0,2)} &=&  -  e_l^4 \, \left[\left(2 \ln{x}\ln{1-x}+\frac{3}{2}\ln{x}\right) p_{qq}(x) + \frac{3+7x}{2}\ln{x} \right. 
\\ \nn &+& \left.  \frac{1+x}{2}{\lnn{x}}+5(1-x) +\left(\frac{\pi^2}{2}-\frac{3}{8}-6 \zeta_3\right) \delta(1-x)  \right] 
\\ &-& e_l^2  \left(\sum_{f}\,e_f^2\right) \left[  \frac{4}{3}(1-x) + p_{qq}(x) \left( \frac{2}{3} \ln{x}+\frac{10}{9}  \right) +\left(\frac{2 \pi^2}{9}+\frac{1}{6}\right) \delta(1-x)\right] \, ,
\\ P_{l\bar{l}}^{V(0,2)} &=& \frac{e_l^4}{e_q^4} \, P_{q\bar{q}}^{V(0,2)} \, ,
\\ P_{lL}^{S(0,2)} &=& P_{l\bar{L}}^{S(0,2)}= e_l^2 \, e_L^2 \, p_s(x) \, . 
\eeqn
Mixed quark-lepton evolution kernels are given by
\beqn 
P_{lq}^{S(0,2)} &=& P_{l\bar{q}}^{S(0,2)}= e_l^2 \, e_q^2 \, p_s(x) \, ,
\\ P_{ql}^{S(0,2)} &=& P_{q\bar{l}}^{S(0,2)}= C_A \, e_l^2 \, e_q^2 \, p_s(x) \, ,
\eeqn
and we notice that they share the same functional dependence, with the exception of the global normalization (influenced by the average over the quantum numbers of the initial particle). Finally, for the photon splitting kernel we have
\beqn
\nn P_{\gamma \gamma}^{(0,2)} &=&  \left( \sum_{f}\, e_f^4\right)  \left[ -16 +8 x +\frac{20}{3}x^2+\frac{4}{3x} -(6+10x) \ln{x} \right.
\\ &-& \left. \vphantom{\frac{4}{3x}} 2(1+x)\lnn{x}  -\delta(1-x) \right] \, , \ \
\eeqn
that, at this order, includes both real and virtual corrections, in contrast with ${\cal O}(\alpha \, \alphas)$ contributions \cite{deFlorian:2015ujt}.

\section{Phenomenological impact of QED corrections}
\label{sec:pheno}
According to the expansion shown in \Eq{eq:expansionkernels}, the weight of higher-order corrections is suppressed by powers of $\alpha$ and $\alphas$. In fact, working at $\mu=M_Z$, we have $a=1.2434 \times 10^{-3}$ and $\as=1.8860 \times 10^{-2}$. Thus, we anticipate that QED contributions to the AP kernels are small compared to pure QCD kernels. However, it might still happen that their effects become magnified due to the specific shape of the different PDFs. For this reason, we perform a study of the QCD and QED contributions to the splitting kernels to anticipate the possible consequences in the evolution of the PDFs.

\begin{figure}[htb]
\begin{center}
\includegraphics[width=0.60\textwidth]{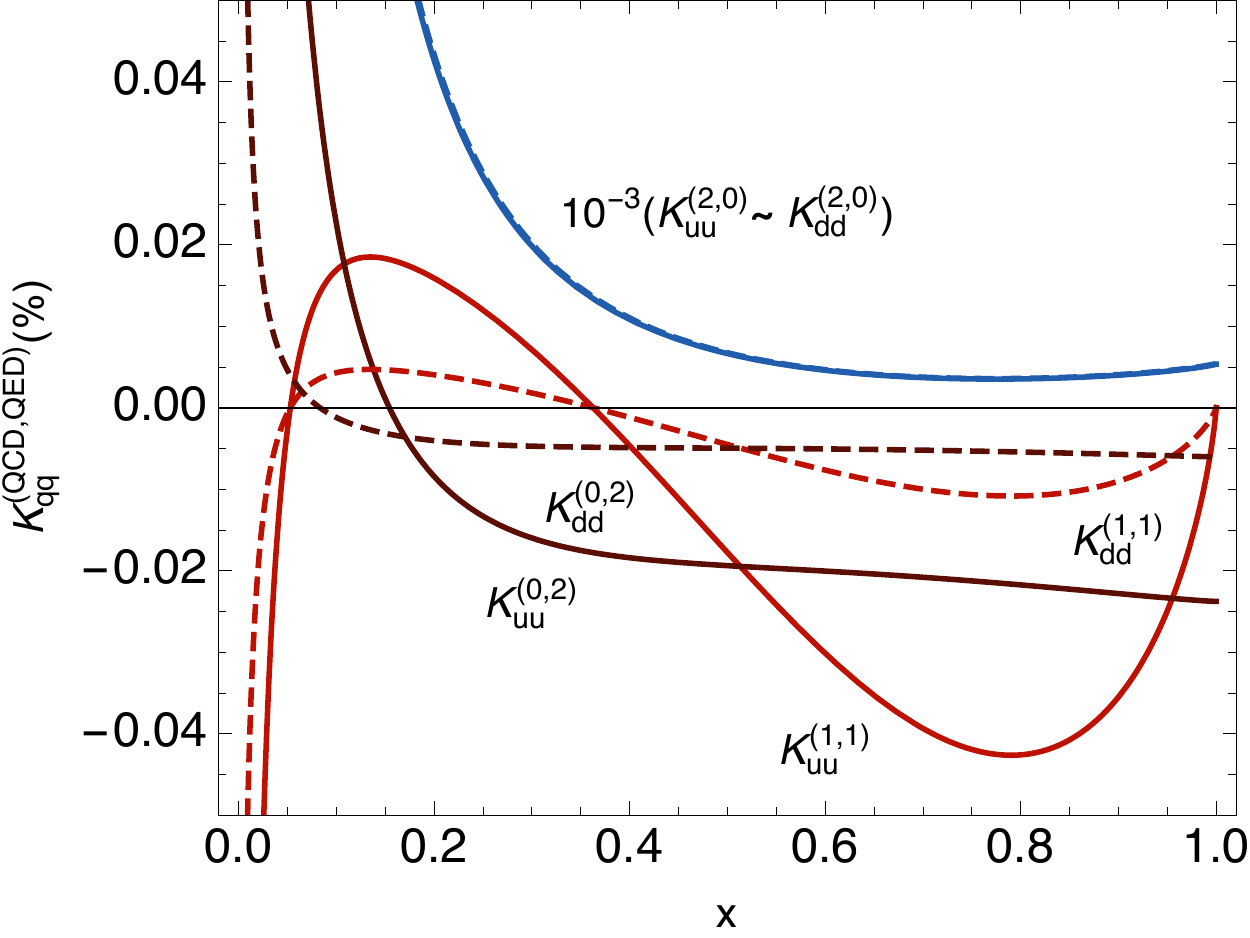}
\caption{$K$ factors for the $qq$ splitting functions ($\%$). We separate among $u$ (solid lines) and $d$ (dashed lines) quarks to study the EM charge effects, and also among the different perturbative orders. Notice that $\alphas^2$ terms are dominant (they are suppressed by a factor $10^{3}$ in this plot) and they exhibit almost the same behaviour for both $u$ and $d$ quarks.} 
\label{fig:PqqCORRECTIONS}
\end{center}
\end{figure}
\begin{figure}[htb]
\begin{center}
\includegraphics[width=0.48\textwidth]{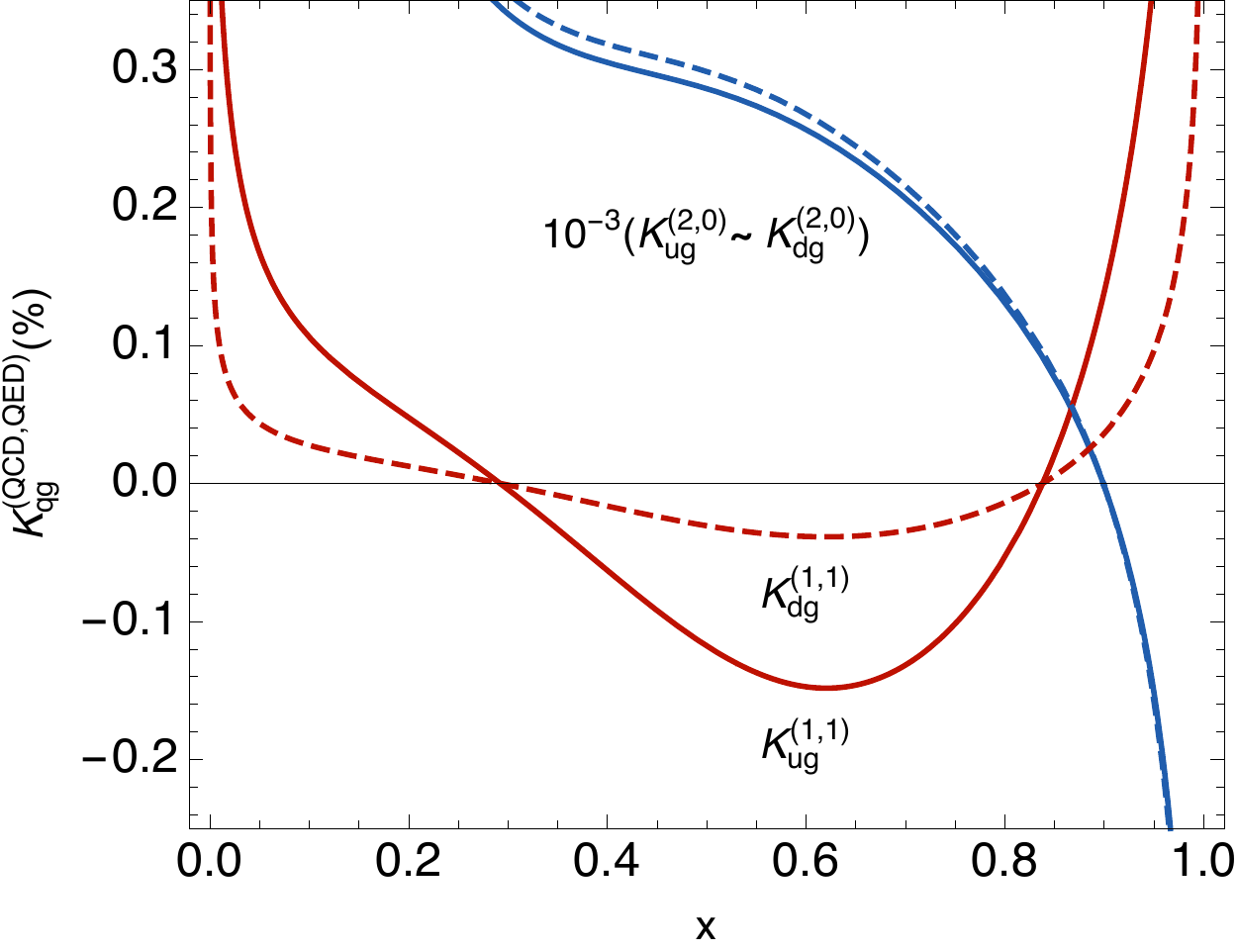} $\quad$
\includegraphics[width=0.48\textwidth]{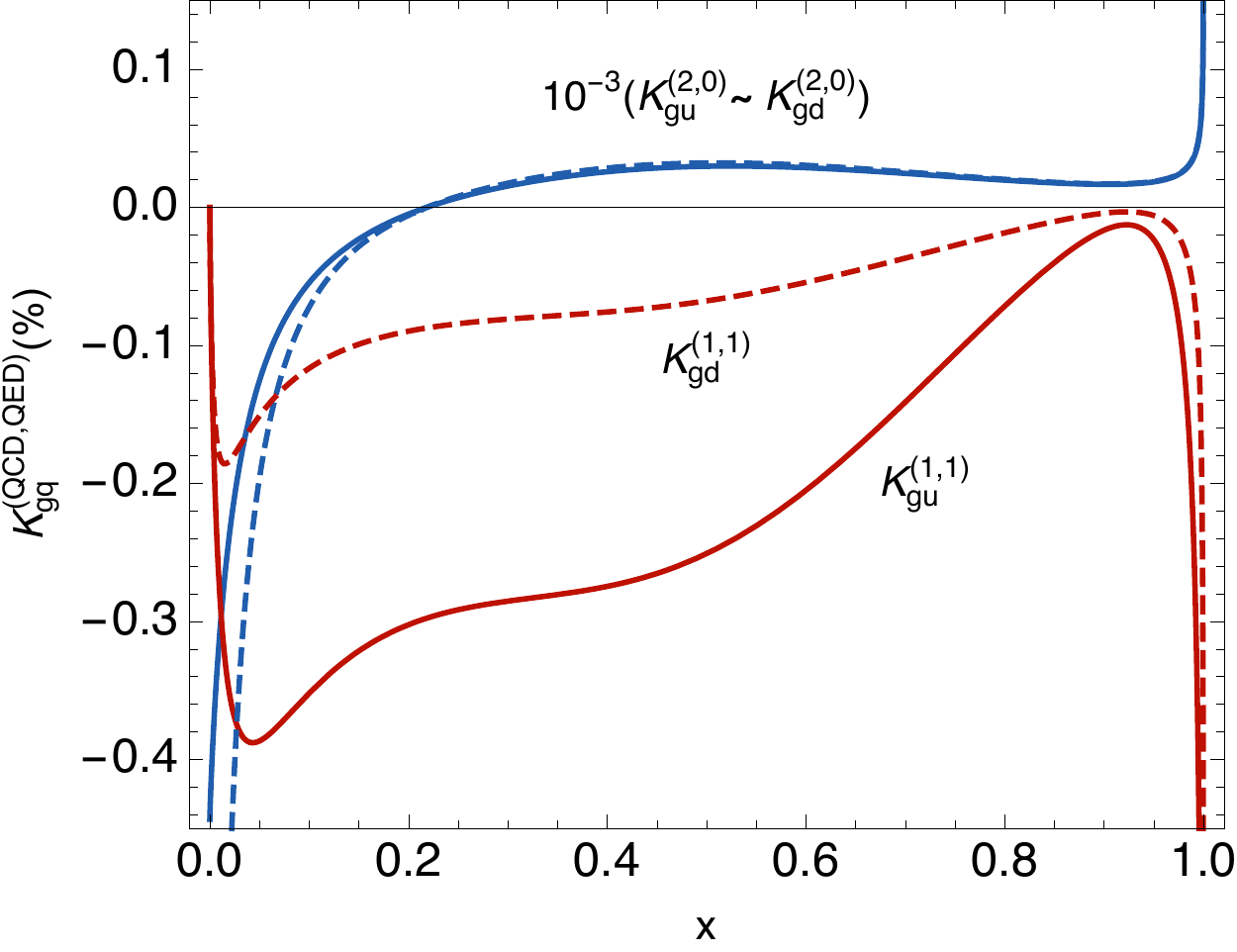} 
\caption{$K$ factors for the $qg$ (left) and $gq$ (right) splitting functions ($\%$). We include ${\cal O}(\alphas^2)$ and ${\cal O}(\alpha \, \alphas)$ contributions, and we also distinguish according to the EM charge of the involved quark; solid lines are used for $u$ quarks, whilst $d$ quarks are displayed with dashed lines. The ${\cal O}(\alphas^2)$ term is dominant, and we suppress it by a factor $10^{3}$ in order to improve the graphical presentation.} 
\label{fig:PqgCORRECTIONS}
\end{center}
\end{figure}

Let's start with the analysis of the pure quark kernels $P_{qq}$. We define the ratio
\beqn
K^{(i,j)}_{ab} &=& \as^i \, a^j \, \frac{P^{(i,j)}_{ab} (x)}{P^{\rm LO}_{ab}(x)} \, ,
\eeqn
where $P^{\rm LO}_{ab}(x)$ is the contribution to the evolution kernel at the lowest order in $\alpha$ and $\alphas$. Notice that 
\beq
\label{eq:tree}
P^{\rm LO}_{ab} = \as \,  P^{(1,0)}_{ab} + a \, P^{(0,1)}_{ab} \, ,
\eeq  
i.e. $P^{\rm LO}_{ab}$ is not necessarily the lowest order contribution in only one of the couplings. We extract the ${\cal O}(\alphas^2)$ contributions from Refs. \cite{Curci:1980uw, Furmanski:1980cm} and the ${\cal O}(\alpha \, \alphas)$ ones from Ref. \cite{deFlorian:2015ujt}; the resulting plot is given in Fig. \ref{fig:PqqCORRECTIONS}. We distinguish there among quarks belonging to the up and down sector, respectively. As expected, deviations arising from QED corrections for $u$ quarks turn out to be bigger than those for $d$ quarks, since they are proportional to $e_q^2$. $P^{(2,0)}_{qq}$ terms are dominant in both cases; they represent a ${\cal O}(10 \%)$ correction, at least. However, the other corrections are of the same order of magnitude; approximately $\pm 0.04 \%$. Except from the singular behaviour in the limit $x \to 0$, there is a positive enhancement of $P^{(1,1)}_{qq}$ for $x \approx 0.10-0.15$ and a negative one for $x \approx 0.75-0.80$.

\begin{figure}[t]
\begin{center}
\includegraphics[width=0.48\textwidth]{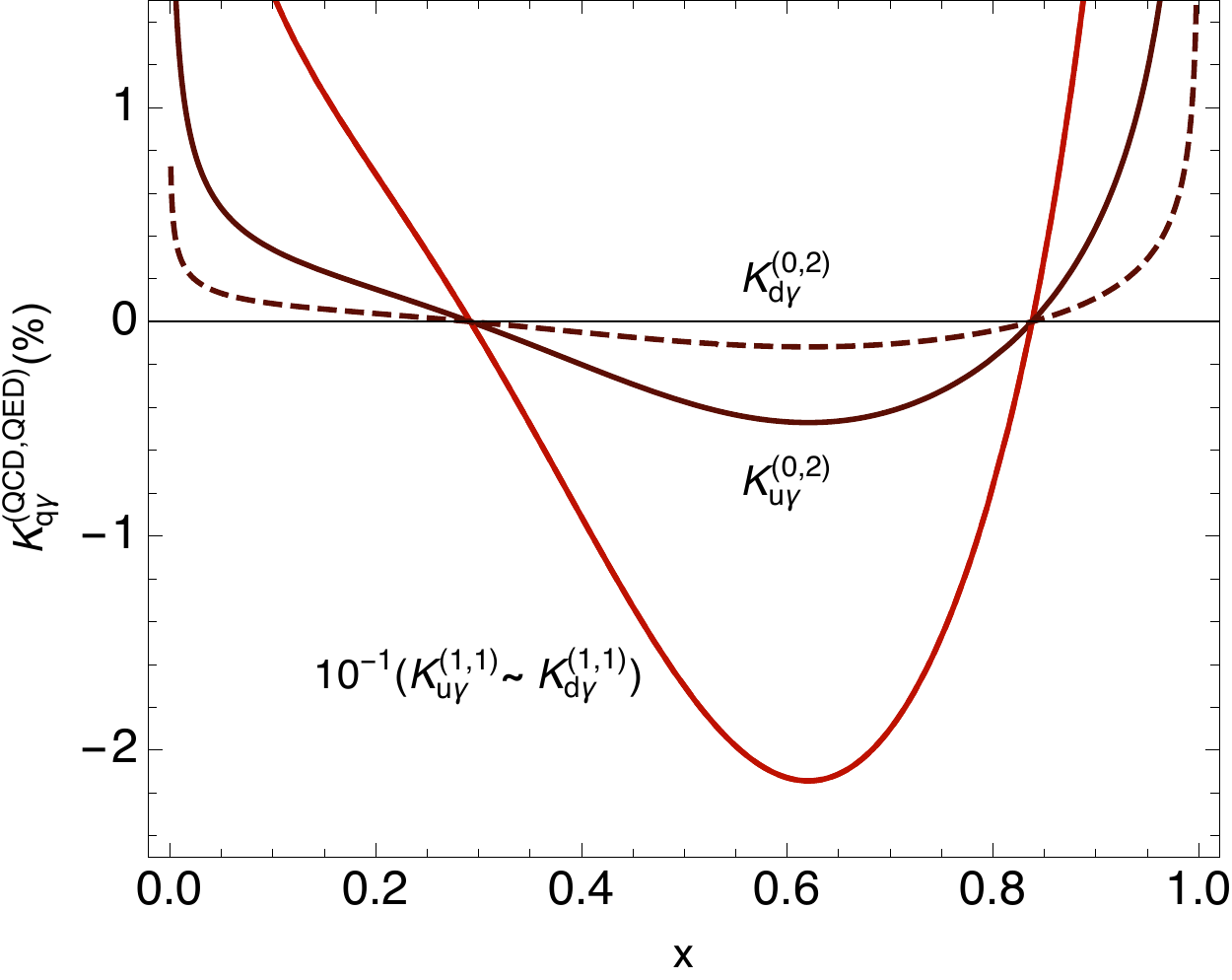} $\quad$
\includegraphics[width=0.48\textwidth]{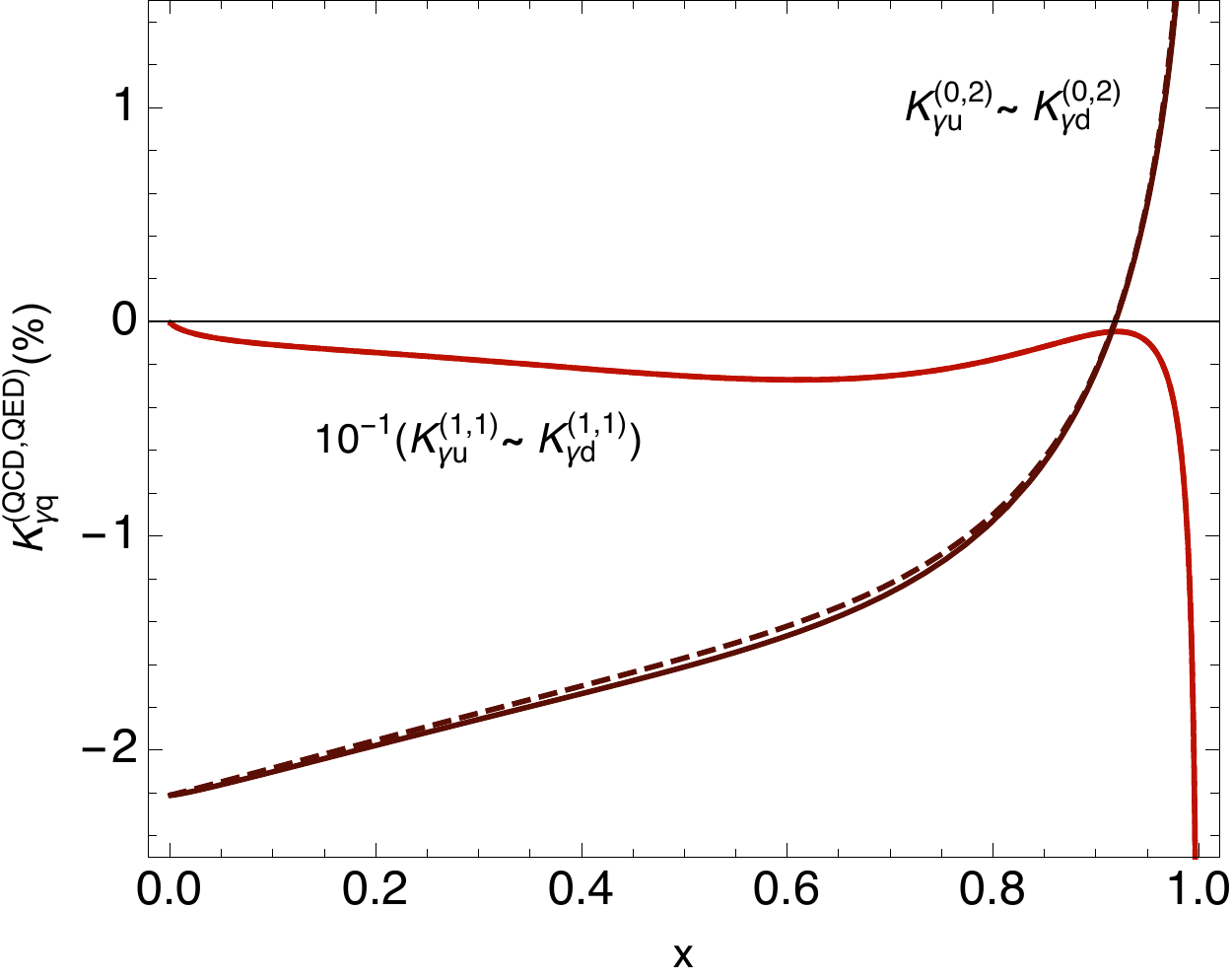} 
\caption{$K$ factors for the $q\gamma$ (left) and $\gamma q$ (right) splitting functions ($\%$). $\alpha^2$ and $\alpha\,  \alphas$ terms are included, with the last one being the dominant contribution. The EM charge distinction is enhanced in $P^{(i,j)}_{q \gamma}$ splitting, around $x \approx 0.65$. Mixed QCD--QED contributions are suppresed by a factor $10$ to improve the visibility in the plot.} 
\label{fig:PqgammaCORRECTIONS}
\end{center}
\end{figure} 
\begin{figure}[htb]
\begin{center}
\includegraphics[width=0.60\textwidth]{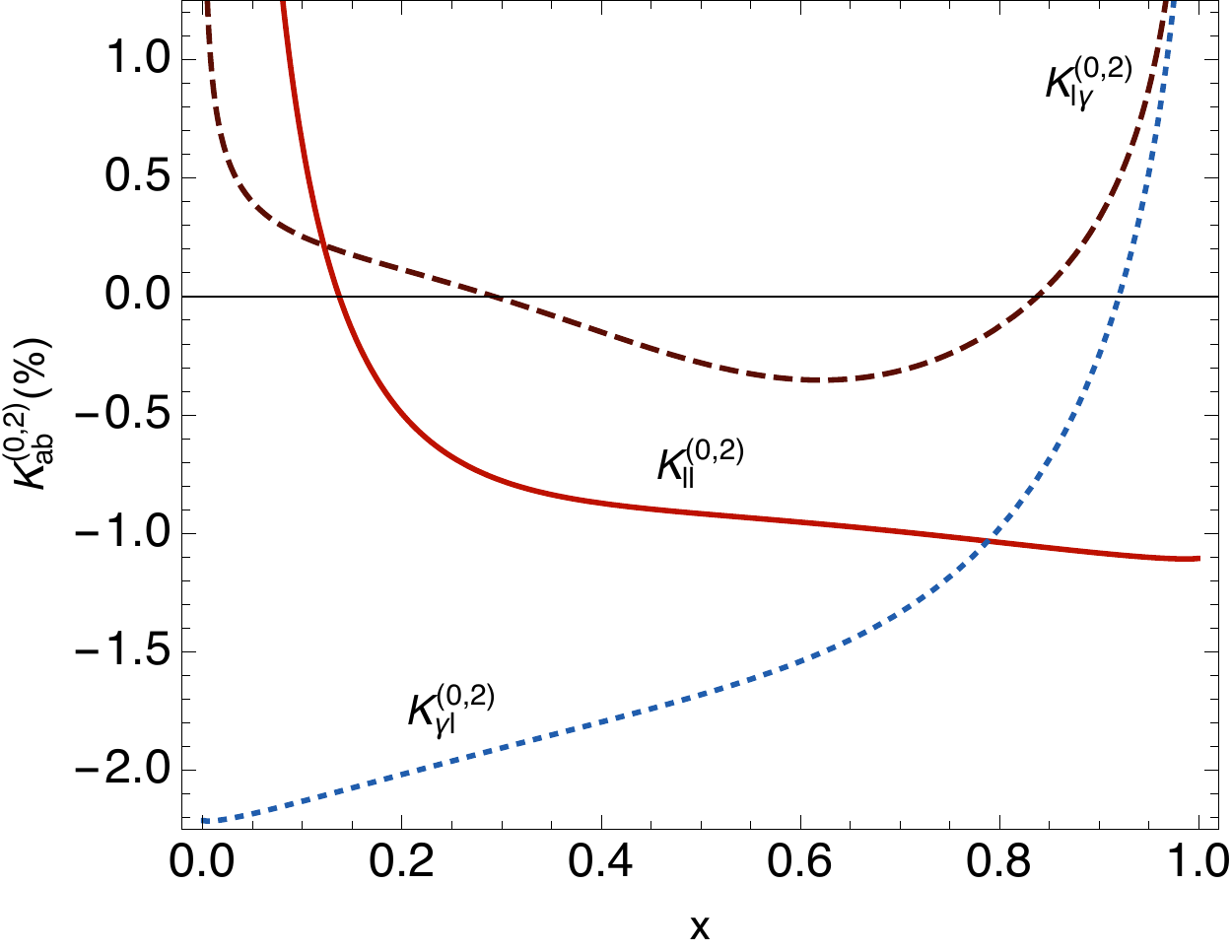}
\caption{$K$ factors for the ${\cal O}(\alpha^2)$ corrections to the splitting functions involving leptons. In the range $0.1 \leq x \leq 0.95$, these contributions represent less than $2 \%$; they become more sizable near $x=0$ (for $K^{(0,2)}_{l\gamma}$ and $K^{(0,2)}_{ll}$) and $x=1$ (for $K^{(0,2)}_{l\gamma}$ and $K^{(0,2)}_{\gamma l}$).} 
\label{fig:PllCORRECTIONS}
\end{center}
\end{figure}

A similar analysis can be performed for $P_{qg}$ and $P_{gq}$ (Fig. \ref{fig:PqgCORRECTIONS}). Pure QCD contributions to the splittings involving gluons are dominant against mixed QCD--QED ones; in any case, these contributions become increasingly relevant in the low $x$ region. Since gluon PDFs are magnified in that region, we expect a non-negligible effect in the evolution. The small EM charge separation observed in the $P_{qg}$ kernel for $\alpha^2$ correction originates from the normalization of the $K$ factor via Eq. (\ref{eq:tree}).

On the other hand, kernels involving a photon receive larger QED corrections, as observed for $P_{q \gamma}$ and $P_{\gamma q}$ (Fig. \ref{fig:PqgammaCORRECTIONS}). Mixed ${\cal O}(\alpha \, \alphas)$ QCD--QED contributions can reach the $20\%$ level for $P_{q \gamma}$, while the two-loop QED terms modify the photon initiated kernel by up to $2\%$ at small $x$. Furthermore, kernels involving leptons provide a non-trivial modification of QCD PDFs at ${\cal O}(\alpha^2)$. In Fig. \ref{fig:PllCORRECTIONS} we plot the $K$ factors for $P_{ll}$, $P_{l \gamma}$ and $P_{\gamma l}$, respectively. Again, corrections reach the $2\%$ level for the photon initiated kernels that can produce non-negligible effects to the photon distribution in a global analysis.

\begin{figure}[htb]
\begin{center}
\includegraphics[width=0.31\textwidth]{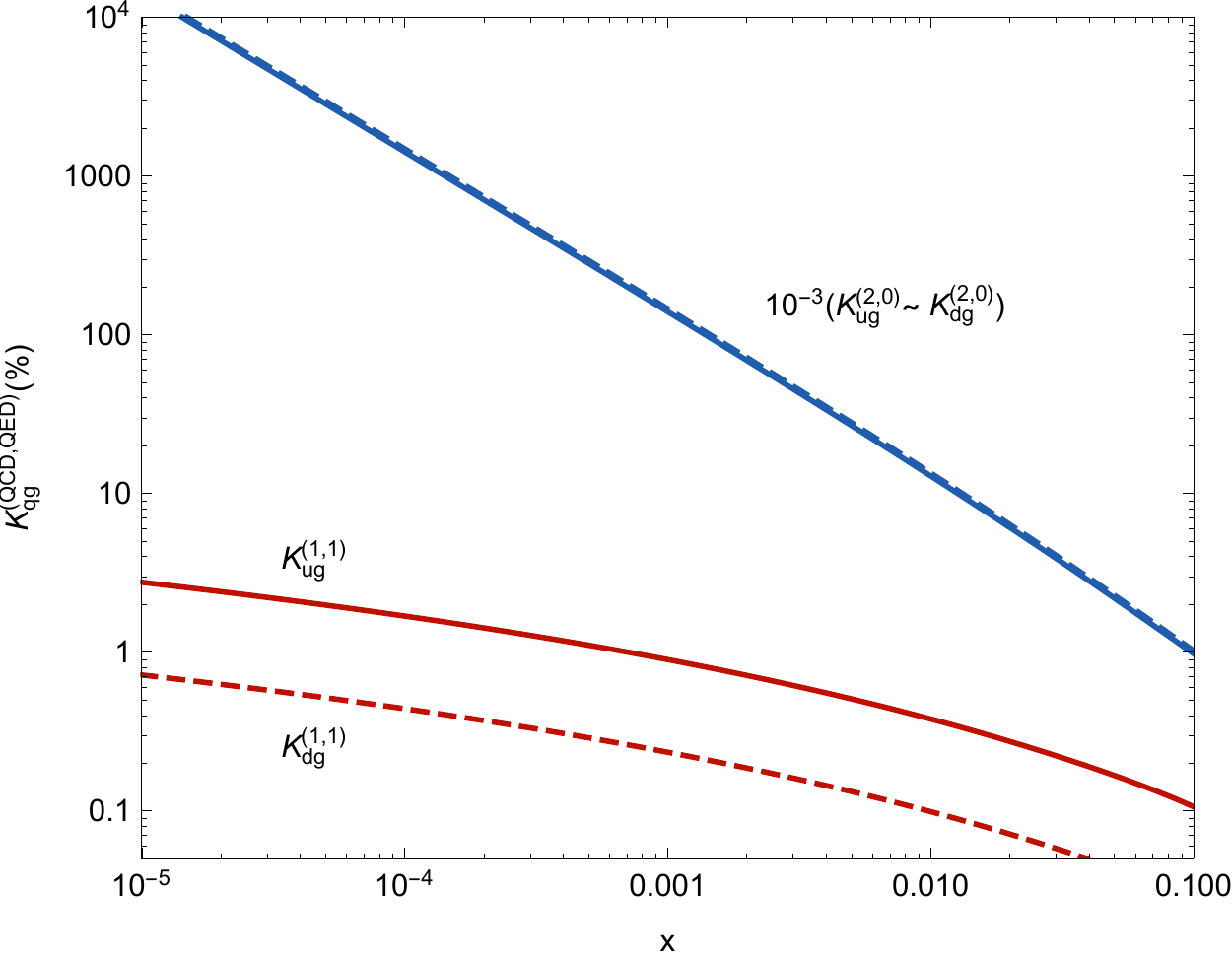} \quad \includegraphics[width=0.31\textwidth]{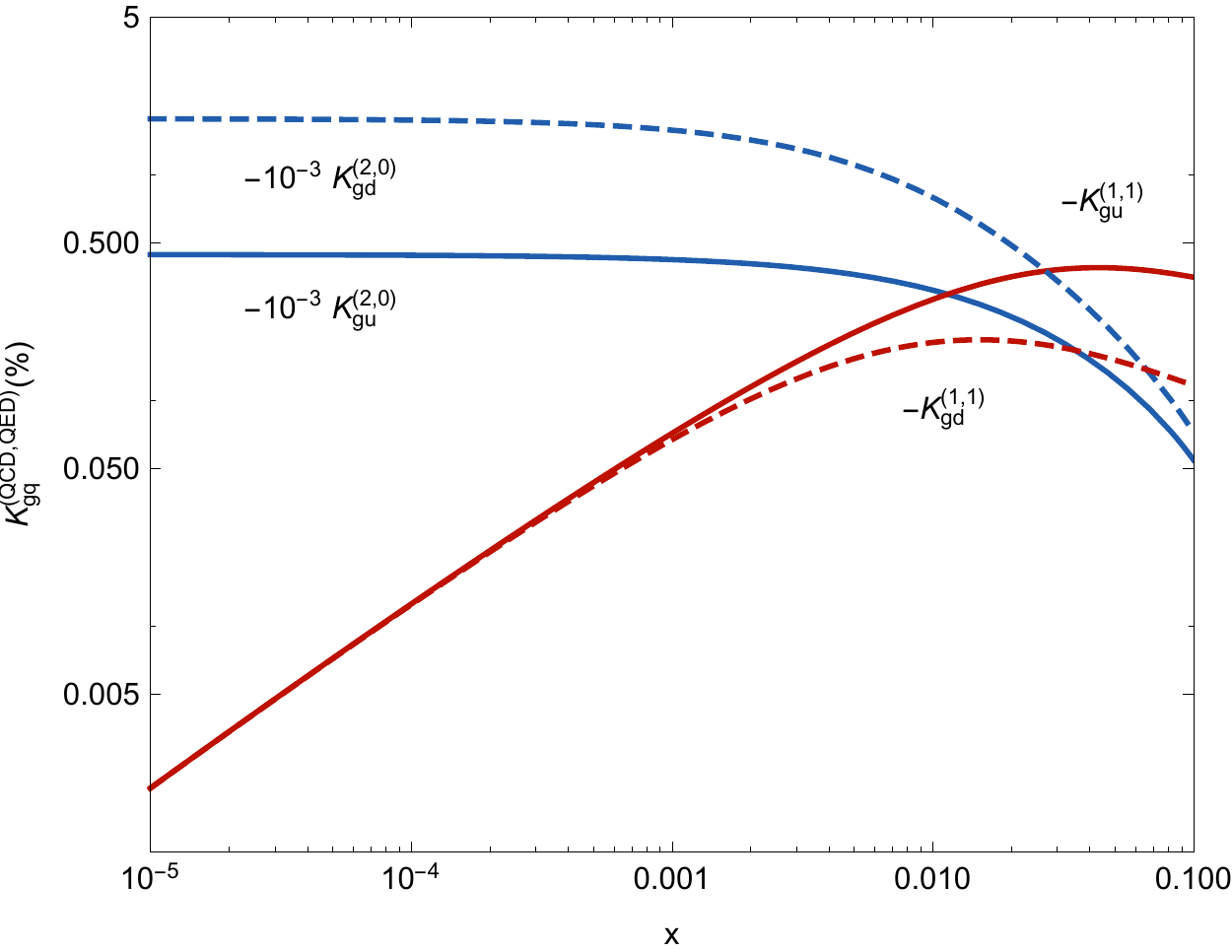} \quad \includegraphics[width=0.30\textwidth]{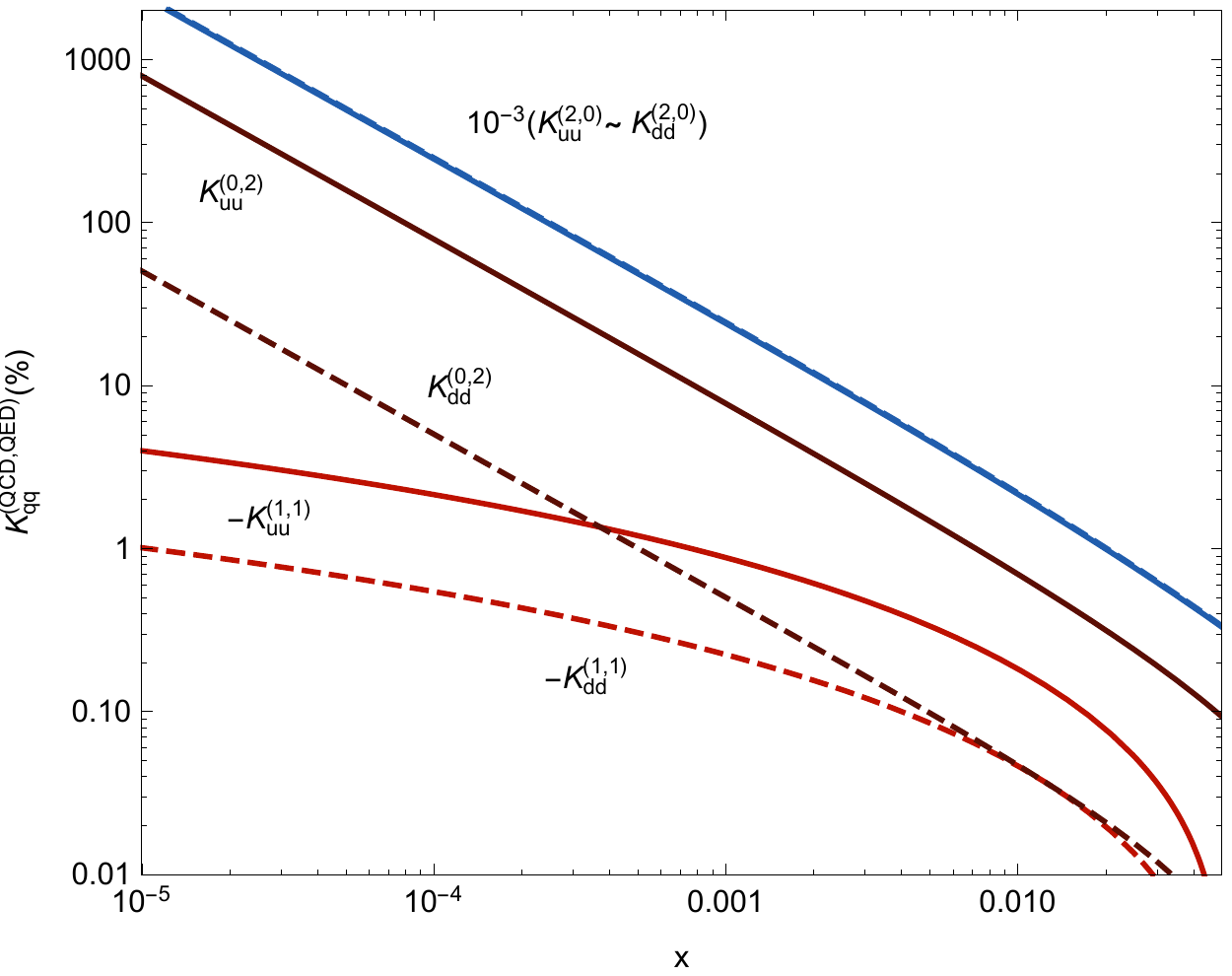}
\includegraphics[width=0.31\textwidth]{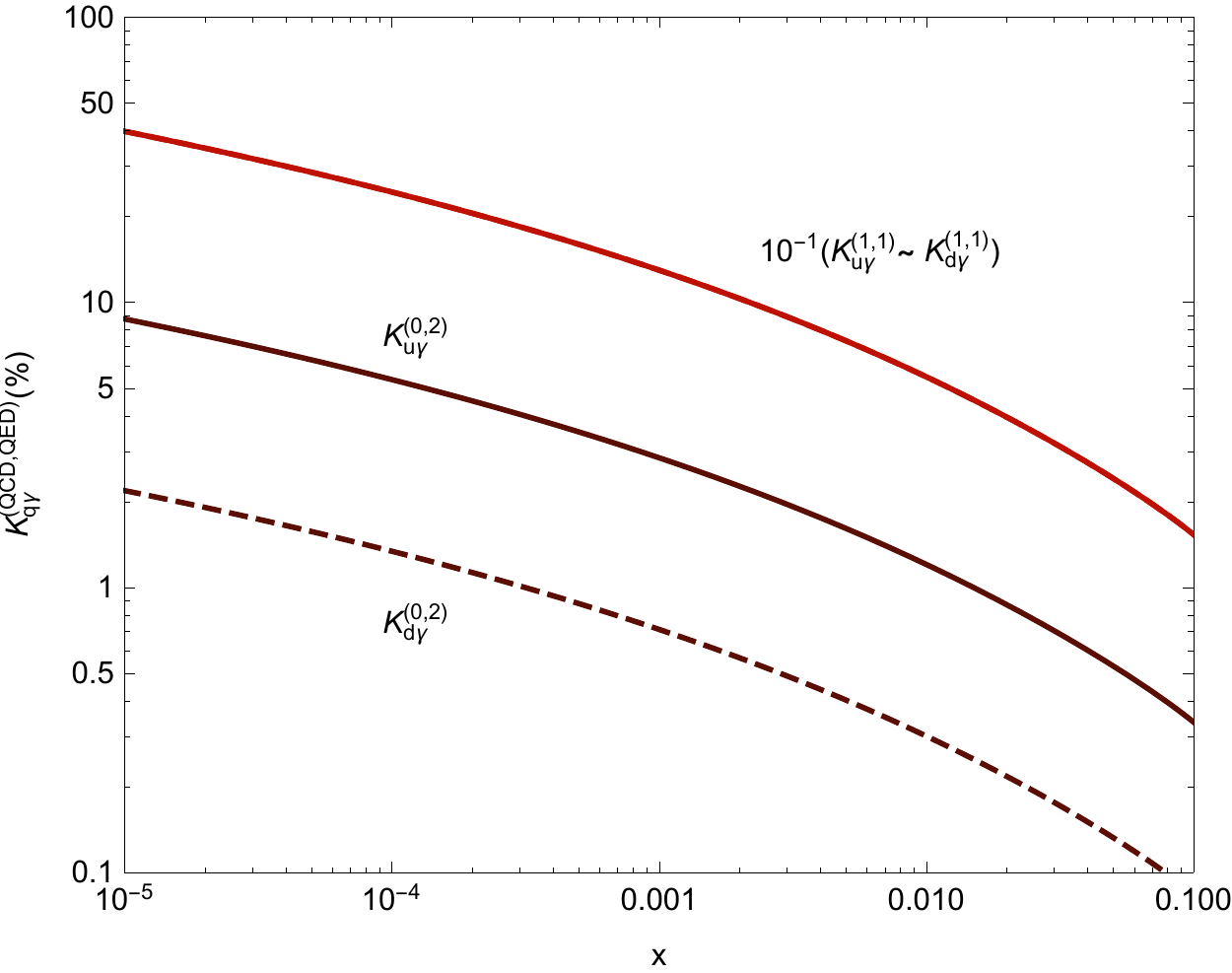} \quad \includegraphics[width=0.315\textwidth]{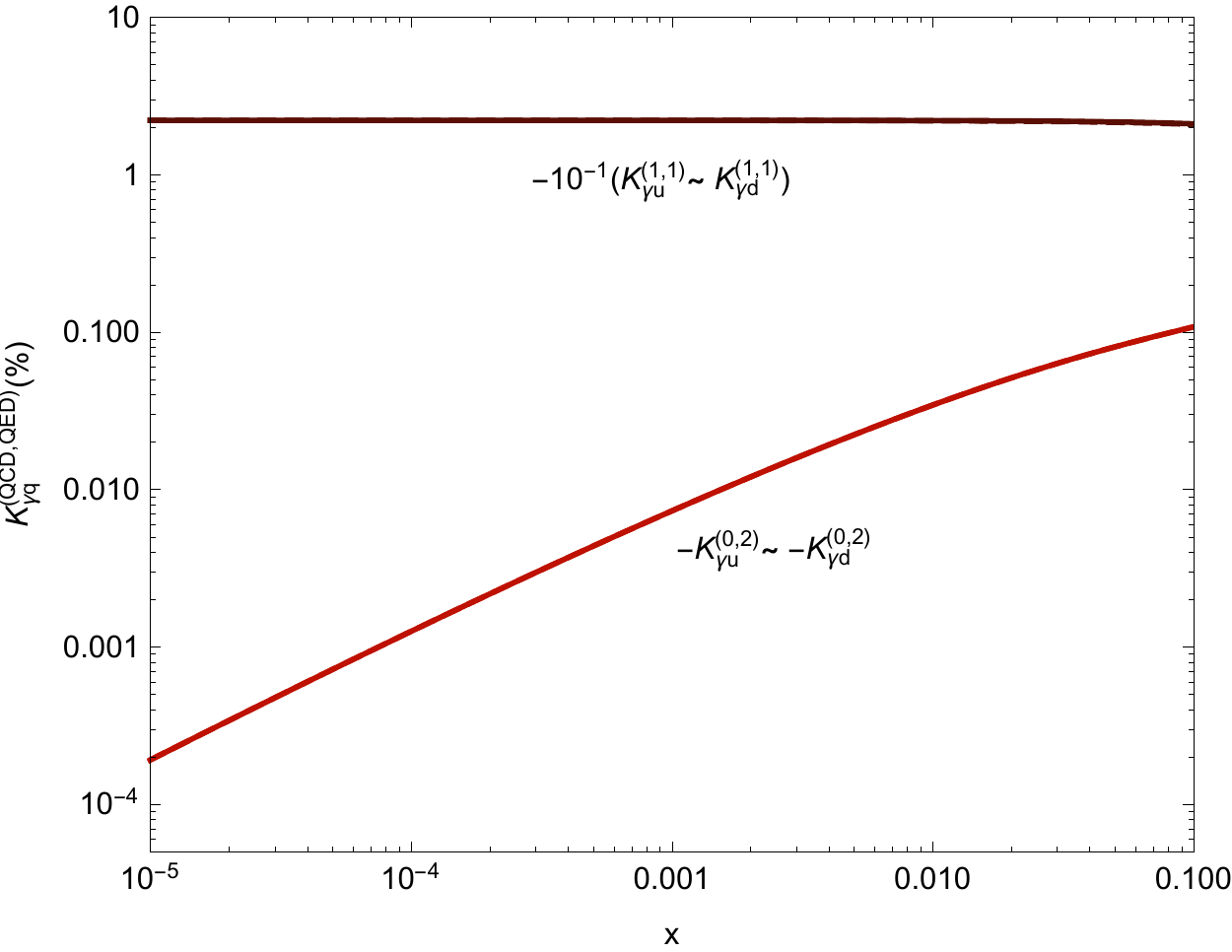} \quad \includegraphics[width=0.31\textwidth]{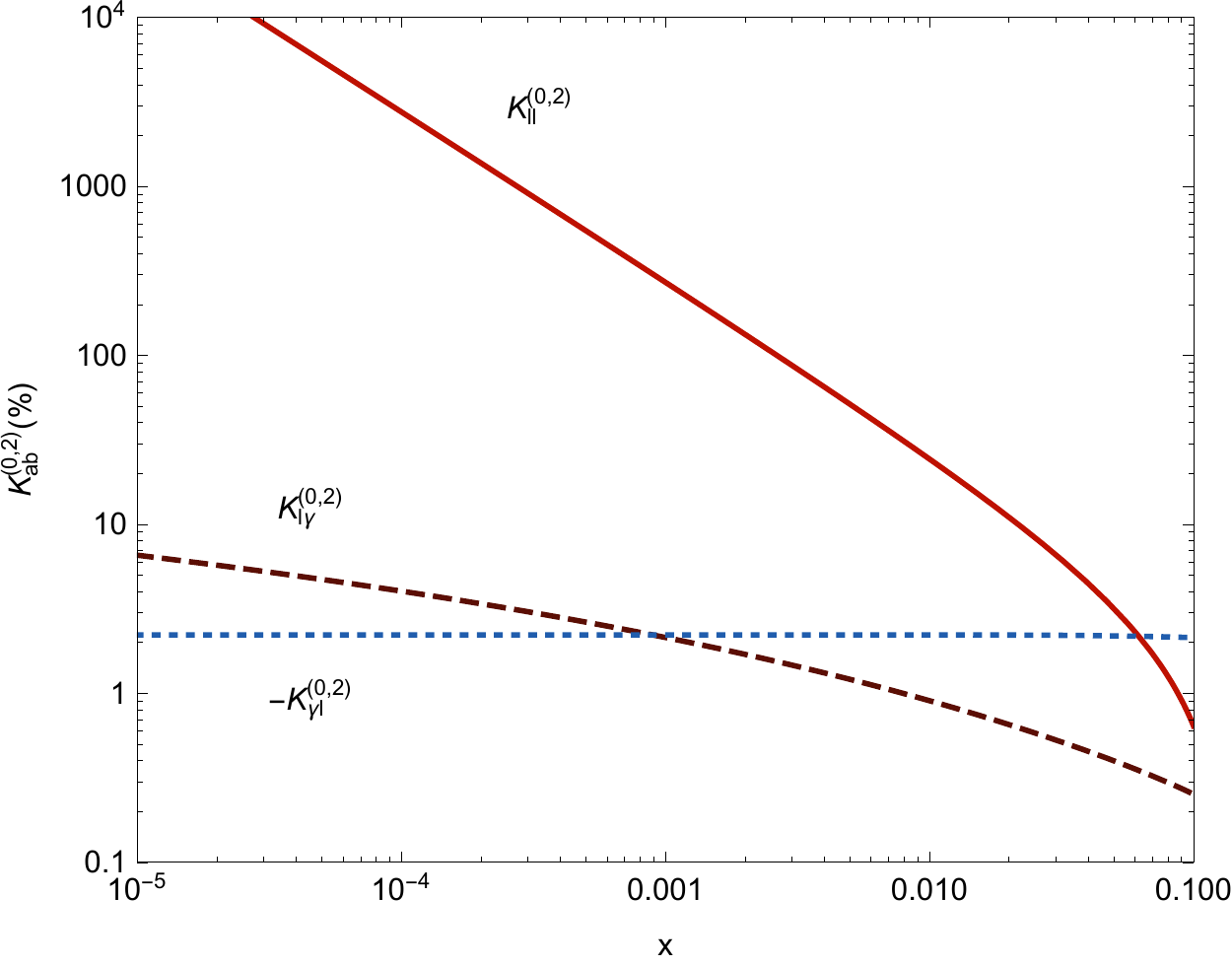}
\caption{$K$ factors in the low-$x$ region. We plot the ${\cal O}(\alphas^2)$, ${\cal O}(\alpha \, \alphas)$ and ${\cal O}(\alpha^2)$ corrections to the mixed QCD-QED splitting functions . In the first line, we considered $K_{qg}$ (left), $K_{gq}$ (center) and $K_{qq}$ (right). In the second one, we show $K_{q \gamma}$ (left), $K_{\gamma q}$ (center) and the ${\cal O}(\alpha^2)$ corrections to the kernels involving leptons (right).} 
\label{fig:plotslogs}
\end{center}
\end{figure}

\section{Conclusions}
\label{sec:conclusions}
In this paper, we have presented for the first time explicit expressions for the Altarelli-Parisi splitting kernels to ${\cal O}(\alpha^2 )$, completing the computation of the two-loop kernels needed to study the evolution of parton distributions to the precision achievable at the LHC. 
The full set of kernels includes those related to both photon and leptonic densities, the latest being allowed to mix in the evolution with parton distributions, mixing that starts at two-loops in QED.

We have obtained the corresponding kernels from the well-known NLO QCD corrections to the splitting functions, after carefully applying a well-defined algorithm to take the Abelian limit of the pure QCD expressions.

Finally, we have performed a phenomenological analysis to study the implications of these corrections in the splitting functions. We find that two-loop corrections are negligible for the pure quark kernels, but become sizable for $P_{qg}$ and $P_{gq}$ at small $x$ values (see Fig. \ref{fig:plotslogs}). The effect of QED corrections turns out to generate ${\cal O}(2\%)$ corrections for the splitting functions initiated by photons, which will alter the shape and size of the photon and leptonic distribution functions in a global analysis.

  \subsection*{Acknowledgments}
This work is partially supported by  CONICET, ANPCyT,
by the Spanish Government and EU ERDF funds
 (grants FPA2014-53631-C2-1-P and SEV-2014-0398) and by GV (PROMETEUII/2013/007).

\end{document}